\documentclass[fleqn,10pt]{wlscirep}
\usepackage[utf8]{inputenc}
\usepackage[T1]{fontenc}
\title{Matters Arising: 
Modeling and observation of mid-infrared nonlocality in effective epsilon-near-zero ultranarrow coaxial apertures
}

\author[1,*]{Reuven Gordon}

\affil[1]{Department of Electrical and Computer Engineering, University of Victoria, Victoria, BC, V8W 2Y2}

\author[1]{Ali Khademi}
\author[1]{Ghazal Hajisalem}
\affil[*]{rgordon@uvic.ca}

\begin{abstract}
In ``Modeling and observation of mid-infrared nonlocality in effective epsilon-near-zero ultranarrow coaxial apertures" the authors perform infrared transmission measurements on coaxial aperture arrays in metal films~\cite{yoo2019modeling}. They claim that the blue-shift of the resonances is the result of nonlocality. They claim that roughness will not produce a shift. This is contrary to past results in plasmonics, to our simulations and to past published claims from some of the same authors. Furthermore, the effect of planarization that occurs for atomic layer deposition, as has been reported elsewhere, will produce a blue-shift. Finally, discrepancies between different nonlocal models, time-dependent density functional theory and other experimental observations all call into question the accuracy of the particular nonlocal model chosen for this regime.   
\end{abstract}
\begin{document}

\flushbottom
\maketitle
%
%

The authors claim that ``
\ldots the presence of roughness would randomly shift the resonance of each aperture toward higher or lower frequencies, producing a global broadening of the resonance without affecting the peak center of mass". The simple ``averaging-out" that the authors are suggesting is not consistent with the literature for surface plasmons. In the 1988 book: ``Surface Plasmons on Smooth and Rough Surfaces and on Gratings" Raether reviews the prior literature on surface plasmons for the case of small roughness and large roughness\cite{raether1988surface}. The result was that for small roughness, the wavevector increases systematically (citing a work by Kretschmann and coworkers~\cite{kretschmann1979splitting} among other works). For larger roughness, there is localization. Since Yoo et al. are attributing the resonance to the wavevector, the resonance should shift systematically with the roughness and not average out. The systematic shift should be a red-shift or a blue-shift, depending on the level of roughness. We argue that the roughness in the experiment is in the blue-shift (large roughness) regime.

\begin{figure}
    \centering
    \includegraphics{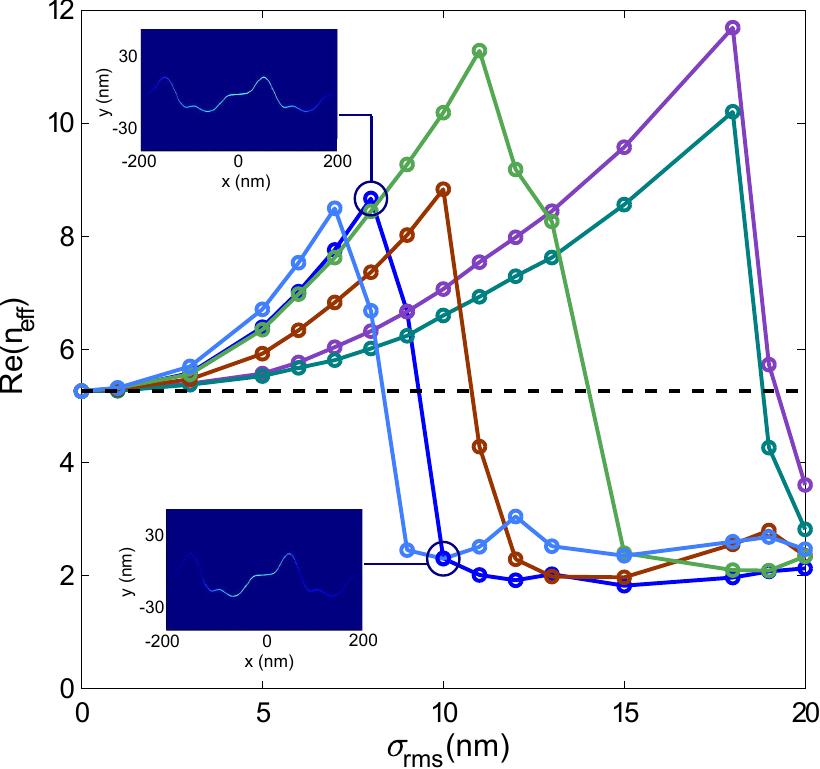}
    \caption{Impact of increasing roughness (standard deviation $\sigma_{\mathrm{rms}}$) on the real part of the effective index for a 400~nm long 1~nm wide alumina slit in gold at 7~$\mu$m free space wavelength. Six different random distributions of roughness are shown. Small roughness uniformly increases the effective index until localization occurs at roughness around 10~nm when it abruptly drops (cut-off). The dashed line shows the real part of the effective index without roughness for comparison. The insets show the mode field intensity for the same roughness pattern but increasing the standard deviation (roughness amplitude). The $\sigma_{\mathrm{rms}}$ = 8~nm field is distributed across the slit and the $\sigma_{\mathrm{rms}}$ =10 nm field is localized. }
    \label{fig:roughness}
\end{figure}

We recognize that the gap plasmon is different from the surface plasmon on a single surface considered in that book. Therefore, we have performed simulations of the propagation in alumina slits in gold while adding roughness, as shown in Figure~\ref{fig:roughness}. Our finding was in line with Raether's book: for small roughness, there is a systematic shift in the wavevector to higher values. For large roughness ($>10$~nm standard deviation for a 1~nm slit), localization sets in and the cut-off occurs for shorter wavelengths. 

We have inspected Fig. 1b in the work of Yoo et al. and we estimate using digital measuring tools that the roughness is about 40~nm peak-to-peak for each of the coaxial apertures shown in that figure. Therefore, we believe that the work is in the regime where roughness can lead to localization and that this effect should be considered as a contributor to the observed blue-shift.  

As a second point related to roughness, several works have reported on the planarizing effect of atomic layer deposition (ALD)~\cite{ritala1999perfectly,fry2017field,lau2014surface}. Planarization increases the size of the gap above the nominal ALD thickness in regions around large features (as compared with the thickness of the deposited layer)~\cite{fry2017field}.
We have shown previously that planarization will result in a perceived blue-shift of the resonances that can be mistaken for nonlocal effects~\cite{hajisalem2014effect}. This was done by comparing as-deposited gold with ultra-flat gold produced by template stripping. By inspection of Figure~1(b) of the paper, the rough features are comparable in size to the gap (and typically larger). Therefore, planarization is a plausible contribution to at least part of the measured blue-shift and it should be quantified.

Finally, it should be considered whether it is suitable to apply nonlocal hydrodynamic theory for the electron response that invokes a local density approximation for the electrons and hard wall boundaries. The failure of these approximations is particularly evident at the boundary of the metal where the electron wavefunction spills out and leads to tunneling for small enough gaps. One can estimate the appropriate length scale for this spill out to be the size of the gap for which there is tunneling. In some works this is between 0.3~nm and 0.5~nm. This a significant contribution that needs to be considered when evaluating gaps that are only 1~nm in size and observing shifts less than 20$\%$. 

Next we consider what other models are available that incorporate nonlocal effects. Time-domain density functional theory (TDDFT) has shown significant differences from the hydrodynamic theory in this regime~\cite{teperik2013robust}: The local Drude model is actually closer to the TDDFT results than the nonlocal model when screening was ignored. Including screening made the local and nonlocal models almost indistinguishable.  

Continuing consideration of the appropriateness of the model chosen, Yoo et al. discuss the hard wall boundary neglecting spill out: ``These effects only become important for gaps below half a nanometer, and can therefore be safely neglected in this work."~\cite{yoo2019modeling} Yet, one of the co-authors, has published an earlier work arguing that for a 2~nm gap, a model that incorporates ``soft" boundaries, the quantum hydrodynamic theory, would produce the \emph{opposite} shift to the model used in this work -- a red-shift instead of a blue-shift~\cite{ciraci2017current}. There appears to be a contradiction about when hard boundaries can be used. (Those earlier calculations were for a simpler metal (Na); however, others have considered similar effects for the noble metal Ag and the nonlocal shifts are still modified significantly for features comparable to a few nanometers~\cite{toscano2015resonance}.) 

There have been several experimental measurements in this regime where the expected blue-shift was simply not seen: Fitting with a local model worked well (that is, better than the nonlocal fits)~\cite{hajisalem2014effect,doyle2017tunable}. 
For even smaller gaps where the quantum corrected model was developed, the permittivity used was local~\cite{esteban2012bridging}. This quantum corrected model has been applied quantitatively to experiments measuring the plasmon shift and the local intensity (through harmonic generation)~\cite{savage2012revealing,hajisalem2014probing}.

Even for exact same coaxial geometry, some of the same authors of the work of Yoo et al. presented earlier experiments on 10~nm, 7~nm, 4~nm and 2~nm gaps where a local model was used to fit~\cite{yoo2016high}. The peak location in that earlier work was exactly fit with a local model down to 7~nm. Also, the location of the peaks for the 2~nm gap and the 4~nm gap have shifted by between 500~nm and 1000~nm from that previous report. 
In that earlier work, the authors attributed the peak shift to roughness and changes from the bulk dielectric: ``For gap sizes below 4 nm, the mismatch increases likely due to fabrication imperfections such as roughness of the interior metal surfaces. In the size regime presented, even subnanometer changes in the gap width can have large effects on the position of resonant peaks. Additionally, the realistic bulk dielectric constant values for an ALD-grown Al$_2$O$_3$ thin film used in simulation can diverge from experimental values for narrower gaps (below 4 nm)"~\cite{yoo2016high}.  Yet in the present work, the authors say that roughness does not play a role. Further, they do not address the issue of changes in the dielectric constant that they raised in that past work. They introduce a nonlocal model to fit the data.  

An even earlier work from some of the same co-authors on the ALD process (although using a different shape) also fit transmission for gaps down below 1~nm with a local model showing good agreement both in the near-IR and sub-THz regimes~\cite{chen2013atomic}. There appears to be a discrepancy between these two earlier works  (from some of the same co-authors) that both used a local theory and the present paper using a nonlocal theory and claiming roughness does not play a role.  

It is important to note that there do exist experiments where nonlocal effects have been observed clearly. For example, there is a nonlocal response for graphene for THz waves~\cite{lundeberg2017tuning}. In that work, the gaps were much larger and the electrons were tightly confined to the plane of graphene, so that the spill out effects discussed above would not be significant. 

In summary, the assertion that roughness will lead to zero net change in the propagation is not consistent with past results. The impact of roughness requires theoretical investigation as a contributor to the observed blue-shift. The effect of planarization of the atomic layer deposition process should also be quantified as it will produce a blue-shift by making the gaps systematically larger. Also, modification of the dielectric constant in small gaps should be considered. Finally, there is a need for further consideration of the validity of the hard wall boundary and the local electron density approximations of the nonlocal model for this regime where electron spill out is expected to reduce the nonlocal effect. Overall, there are many possible contributors to the observed blue-shift, separate from nonlocal effects.

\section{Acknowledgements}

The author would like to thank Rog\'erio de Sousa for valuable discussions on nonlocal hydrodynamic theory. 

\bibliography{main.bib}

\end{document}